\documentclass[amstex,12pt]{article}     
\usepackage{amssymb,amsmath}           

\newlength{\dinwidth}
\newlength{\dinmargin}
\newtheorem{theorem}{Theorem}[section]
\newtheorem{prop}[theorem]{Proposition}
\newtheorem{lemma}[theorem]{Lemma}

\newenvironment{proof}{\medskip \noindent 
            {\bf Proof.}}{ \hfill $\square$ \medskip}

\newcommand{\ie}{{\it i.e.\ }}

\newcommand{\cf}{{\it cf.\ }}

\newcommand{\nind}{\noindent}
\newcommand{\bmu}{{\mbox{\boldmath $\mu$}}}

\newcommand{\snu}{{\mbox{\footnotesize $\nu$}}}

\newcommand{\sbbeta}{{\mbox{\footnotesize \boldmath $\beta$}}}
\newcommand{\sbmu}{{\mbox{\footnotesize \boldmath $\mu$}}}
\newcommand{\sbnu}{{\mbox{\footnotesize \boldmath $\nu$}}}

\newcommand{\RR}{\mathbb R}
\newcommand{\CC}{\mathbb C}
\newcommand{\NN}{\mathbb N}
\newcommand{\ZZ}{\mathbb Z}

\newcommand{\CD}{{\cal C}_\Delt}
\newcommand{\CU}{{\cal C}}

\newcommand{\sbk}{{\mbox{\footnotesize \boldmath $k$}}}
\newcommand{\be}{{\mbox{\boldmath $e$}}}
\newcommand{\sbe}{{\mbox{\footnotesize \boldmath $e$}}}
\newcommand{\bk}{{\mbox{\boldmath $k$}}}
\newcommand{\bl}{{\mbox{\boldmath $l$}}}

\newcommand{\bp}{{\mbox{\boldmath $p$}}}
\newcommand{\sbp}{{\mbox{\footnotesize \boldmath $p$}}}
\newcommand{\bx}{{\mbox{\boldmath $x$}}}

\newcommand{\Delt}{{B}}

\begin{document}
\title {On Hot Bangs and the Arrow of Time in \\
Relativistic Quantum Field Theory} 
\author{Detlev Buchholz\,  \\[3mm]
Institut f\"ur Theoretische Physik, Universit\"at G\"ottingen\\
37073 G\"ottingen, Germany}
\date{\small Dedicated to Rudolf Haag on the occasion of his 80th birthday}
\maketitle 
\begin{abstract}{\noindent 
A recently proposed method for the characterization
and analysis of local equilibrium states in relativistic quantum
field theory is applied to a simple model. Within this model
states are identified which are locally (but not globally)
in thermal equilibrium and it is shown that their local thermal 
properties evolve according to macroscopic 
equations. The largest space--time regions 
in which local equilibrium states can exist are 
timelike cones. Thus, although the model does 
not describe dissipative effects, such states fix in
a natural manner a time direction. Moreover, generically 
they determine a distinguished space--time point where 
a singularity in the temperature (a hot bang) must have 
occurred if local equilibrium prevailed thereafter. The  
results illustrate how the breaking of the time reflection
symmetry at macroscopic scales manifests itself in a 
microscopic setting.}
\end{abstract}
\section{Introduction} 
\setcounter{equation}{0}
Within the framework of relativistic quantum field theory,
a general method has been established in \cite{BuOjRo} for the 
characterization and analysis of states which are locally 
close to thermal equilibrium. This novel approach is based 
on the idea of comparing states locally with the 
members of a family of thermal reference states consisting
of mixtures of global equilibrium states: If a given state 
happens to coincide at a space--time point 
$x$ with some such reference state to some degree of accuracy, 
\ie if the expectation values of a sufficiently large number 
of distinguished local observables at $x$ coincide in the two 
states, the given state is said to be locally in 
equilibrium at $x$. One can then determine the thermal 
properties of the state at $x$ by noticing that the local
observables have an unambiguous macroscopic interpretation in 
all global equilibrium states. Moreover, the space--time 
evolution of these thermal properties is directly linked to the 
microscopic dynamics.

The application of this method to concrete models thus 
requires the following steps: (1) Determine the global 
equilibrium states (KMS states) of the theory; their mixtures 
define the convex set of thermal reference states. (2)  
Select for each space--time point $x$ a set of 
suitable local observables which are sensitive to thermal
properties. As was explained in \cite{BuOjRo}, a natural
choice consists of the basic observable point fields of the 
theory as well as of density like quantities which are obtained 
by taking normal products. (3) Determine the thermal 
interpretation of these local observables. This is 
accomplished by computing their expectation values in the  
global equilibrium states. The result is a family of 
thermal functions depending on the parameters characterizing
the thermal states, such as temperature and chemical potential;
they describe certain specific intensive macroscopic 
properties of 
the global equilibrium states. (4) After these preparations, 
one can characterize the local equilibrium states in a given
space--time region as follows: at each point of the region,
the mean values of the distinguished local observables 
taken in such a state coincide
with those in some thermal reference state. This 
reference state may vary from point to point. One    
then ascribes to the local equilibrium state 
those macroscopic thermal properties of the respective
reference states which are fixed by the distinguished 
thermal functions. The space--time evolution of these properties can  
be derived directly from the underlying microscopic dynamics.

In the present investigation we apply this scheme to the 
theory of a free massless scalar field. In view of the 
simplicity of this model, it may be necessary to comment 
on the physical significance of such an analysis. It is 
clear from the outset that a non--equilibrium state in this model  
will never approach (local) equilibrium because of the lack 
of interaction. So we will have nothing to say about 
this intriguing problem. But, appealing to 
physics, one knows that, once local equilibrium has been 
reached in a system, there appear in general long--living 
macroscopic structures, such as hydrodynamical flows, where 
dissipative effects are of minor importance. Moreover, if the 
local temperature of a system is sufficiently high 
and the interaction is asymptotically free, these dissipative 
effects 
ought to be small also on dynamical grounds. Thus one may 
expect that massless free field theory provides some 
physically meaningful qualitative information about the 
possible space--time patterns of local equilibrium states 
in hot relativistic systems. 

The present investigation is primarily intended  
as an illustration of the general ideas expounded 
in \cite{BuOjRo}, but it offers some surprises.
First, we will see that in all local equilibrium states 
of the model one can determine locally 
macroscopic properties, such as the particle density
(distribution function), without having to rely on 
an {\it ad hoc} procedure of coarse graining. Second, 
we will establish evolution equations for these macroscopic 
observables, akin to transport equations. 
These equations, together with further constraints 
arising from the statistical interpretation of
the microscopic theory, imply that the largest space--time
regions in which local equilibrium states (not being in 
global equilibrium) can exist have the form of timelike cones. 
Phrased differently, the time reflection symmetry is 
broken in these states, so one can fix 
with their help a time direction. Finally, it turns out
that such states determine generically a unique space--time point
where a \mbox{singularity} in the temperature (a hot bang)
must have occurred if the state was in local equilibrium
thereafter. An example of such a state was already 
presented in \cite{BuOjRo}. The present analysis shows
that the appearance of such singularities is a generic feature
of the model. Incidentally, it provides a consistent 
qualitative picture of the space--time patterns of the 
cosmic background radiation.

The general mathematical framework and scheme 
for the identification and analysis of local equilibrium 
states is outlined in the subsequent section. For a 
more detailed discussion and physical justification of the 
method we refer to \cite{BuOjRo}. In
Sec.\ 3 we describe the model and carry out the preparatory 
steps, indicated above, which are required for the 
analysis of the local equilibrium states. 
Section 4 contains the discussion of the space--time 
behaviour of thermal properties of these  
states and the derivation of macroscopic equations
governing their evolution. In Sec.\ 5 we show that 
local (in contrast to global) equilibrium cannot exist 
in all of space--time and we determine the maximal regions 
admitting this feature. The paper concludes with a 
discussion of these results and some interesting perspectives.
 
\section{General framework and scheme}
\setcounter{equation}{0}

The basic objects in our analysis are local observable 
fields, such as conserved currents and the stress energy tensor, 
denoted generically by $\phi (x)$ (with tensor indices omitted). 
Mathematically, these fields may be interpreted in different
ways. Usually, they are regarded as operator valued 
distributions, \ie their space--time averages with test 
functions $f$ having compact support in Minkowski space $\RR^4$,
\begin{equation} \label{e.smear}
\phi (f) \doteq \int \! dx \, f(x) \phi (x), 
\end{equation}
are assumed to be defined on some common dense and stable 
domain $\cal{D}$ in the underlying Hilbert space $\cal{H}$. 
The finite sums and products of the averaged field operators 
generate a *-algebra $\cal{A}$, called the algebra of 
observables, consisting of all polynomials $A$ of the form  
\begin{equation} \label{polynom}
A=\sum \phi (f_{1}) \phi (f_{2})\cdots \phi (f_{n}). 
\end{equation}
We include also all multiples of the unit operator $1$ 
in ${\cal A}$. As the fields $\phi$ are real (being observable), the  
*-operation (Hermitian conjugation) on $\cal{A}$ is defined by 
\begin{equation} \label{adjoint}
A^{\ast } \doteq \sum \phi(\overline{f_{n}}) \cdots 
\phi (\overline{f_{2}}) \phi (\overline{f_{1}}),
\end{equation}
where $\overline{f}$ denotes the complex conjugate of $f$. 
The Poincar\'e transformations $\lambda \in {\cal P}_+^\uparrow$ 
act on $\cal{A}$ by automorphisms $\alpha_\lambda$,  
\begin{equation} \label{e.transl}
\alpha_\lambda (A) \doteq
\sum \phi (f_{1,\lambda}) \phi (f_{2,\lambda})\cdots \phi (f_{n,\lambda}), 
\end{equation}
where $f_\lambda$ is defined by 
$f_\lambda (x) \doteq D(\lambda ) f(\lambda^{-1} x)$ 
and $D$ is a matrix
representation of ${\cal P}_+^\uparrow$ corresponding to the tensor
character of the respective field $\phi$. If $\lambda$ is a pure
translation, $\lambda = (1,a)$, $a \in \RR^4$, 
we denote the corresponding automorphism
by $\alpha_a$. 
  
The states of the physical systems are described by 
positive normalized expectation
functionals on the algebra $\cal{A}$, generically denoted by $\omega$. 
We recall that these functionals have the defining properties 
$\omega (c_{1}A_{1}+c_{2}A_{2}) = c_{1} \, \omega(A_{1}) +c_{2} \, 
\omega(A_{2}) $, $c_1,c_2 \in \CC$  (linearity),  
$\omega(A^*A) \geq 0$ (positivity), and  
$\omega(1) =1$ (normalization). 
As has been explained in the introduction, an important 
ingredient in the present investigation 
are the global equilibrium states which are characterized by 
the KMS condition \cite{HaHuWi,BrRo}. Given a Lorentz frame, fixed by 
some positive time-like vector $e \in V_+$ of unit length, 
$e^2 = 1$, this condition can be stated as follows.

\nind {\bf KMS condition:} A state $\omega_\beta$ satisfies 
the KMS condition at inverse temperature $\beta > 0$ in the 
given Lorentz frame if for each pair of operators
$A,A^\prime \in {\cal A}$ there is some function $h$ which is analytic  
in the strip $S_\beta \doteq \{ z \in \CC : 0 <
\mbox{Im} z < \beta \} $ 
and continuous at the boundaries such that 
\begin{equation}
h(t) = \omega_\beta(A^\prime \alpha_{te}(A)), 
\ \ h(t+i\beta) = \omega_\beta(\alpha_{te}(A) A^\prime), 
\quad t \in \RR.
\end{equation}
In this situation $\omega_\beta$ is called a KMS state. \\[2mm]
\indent Any KMS state 
$\omega_\beta$ describes an ensemble which is in thermal equilibrium
in the distinguished Lorentz frame, describing the rest system 
of the state. As we have to keep track of both, temperatures 
and rest systems, we combine this information into a four vector 
$\beta e \in V_+$, again denoted by $\beta$. Moreover,
in order to simplify the subsequent discussion, we assume that 
for any given $\beta \in V_+$ the corresponding KMS state 
$\omega_\beta$ is unique. As a consequence of this
uniqueness assumption, the KMS states transform under Poincar\'e
transformations according to 
\begin{equation} \label{transformation}
\omega_\beta \circ \alpha_\lambda^{-1} = \omega_{\Lambda \beta},
\end{equation}
\ie they are isotropic in their rest systems and invariant 
under space-time translations. 
Next, let us introduce the thermal reference states. They are 
defined, for any compact subset $\Delt \subset V_+$, as  
arbitrary mixtures of KMS states $\omega_\beta$ 
with $\beta \in \Delt$. We emphasize that the restriction to 
mixtures with compact temperature support $\Delt$ is merely 
a matter of mathematical convenience in the present context. 
In general one may admit 
larger families of reference states with non--compact
temperature support. In order to obtain a more 
explicit description of the reference states and to avoid 
mathematical subtleties, we assume that the KMS states $\omega_\beta$
are weakly continuous in $\beta$, \ie all functions
\begin{equation} \label{continuity}
\beta \mapsto \omega_\beta(A), \quad A \in {\cal A},
\end{equation}
are continuous. Apart from phase transition points (which 
are excluded here by our uniqueness assumption), this property 
is expected to hold quite generally. With this input, the states 
$\omega_\Delt \in \CD$ can be represented in the form 
\begin{equation} \label{mixture}
\omega_\Delt(A) = 
\int \! d\rho(\beta) \, \omega_\beta(A), \quad A \in {\cal A},
\end{equation}
where $\rho$ is a positive normalized measure which has support 
in $\Delt$. The convex set of all these states will be denoted
by $\CD$ and the elements of $\CU \doteq \bigcup \, \CD$, 
where the union extends over all compact subsets $\Delt \subset V_+$, 
will be our reference states for the characterization and analysis
of the local equilibrium states.

It is mathematically meaningful and conceptually convenient 
in the present analysis 
to consider also the unregularized fields  $\phi (x)$. Because of
their singular nature they have to be interpreted in the form
sense, however. Let us illustrate this for the KMS states. 
In view of the invariance of these states under
space--time translations, the expectation values 
$\omega_\beta (\phi (f))$ do not depend on the choice of 
$f$ as long as the value of the space--time integral of the
test function is kept fixed. One may thus extend the thermal reference 
states to the unsmeared fields $\phi(x)$ by the formula
\begin{equation} \label{expect}
\omega_\Delt (\phi(x)) \doteq \omega_\Delt (\phi(f)),
\end{equation}
where the space--time integral of $f$ is put equal to $1$.
This expression does not depend on $x$. We denote the linear 
space generated by all fields $\phi(x)$ at $x$ by ${\cal Q}_{\, x}$ 
and assume without further mentioning that all states
$\omega$ of interest here can be extended to the spaces ${\cal Q}_{\, x}$ 
for $x$ varying in some region. 

It is crucial in the present approach that the local fields $\phi(x)$ 
provide the same information about properties 
of the thermal reference states as certain macroscopic (central) 
observables. To verify this, we pick a test function $f$
of compact support which integrates to $1$ and put 
$f_n(x) \doteq n^{-4} \, f( n^{-1}x - x_n)$, $n \in \NN$, where $x_n$
is a sequence of translations tending sufficiently
rapidly to spacelike infinity such that the support of $f_n$ lies
in the causal complement of any given bounded region for 
almost all $n$. It follows from the uniqueness of 
the KMS states and their invariance under space--time translations 
by an application of the mean ergodic theorem that in the limit of large $n$ 
\begin{equation} \label{central}
\omega_\beta (\phi(f_n)^*\phi(f_n) )
- \omega_\beta (\phi(f_n)^*) \, \omega_\beta (\phi(f_n)) \rightarrow 0.
\end{equation}
Moreover, because of the local commutativity of observables and
the support properties of the test functions,
$\phi(f_n)$ commutes for almost all $n$ with any given
$A \in {\cal A}$, \ie the operators $\phi(f_n)$ form a central 
sequence in ${\cal A}$. Applying standard arguments, it follows
that the limit 
\begin{equation} \label{phithermal}
     \Phi \doteq  \lim_{n \rightarrow \infty} \, \phi(f_n)
\end{equation}
exists in all thermal reference states and defines a macroscopic observable
commuting with all elements $A \in {\cal A}$ and 
the translations. In fact, making use of (\ref{mixture}),
(\ref{central}) and applying the dominated
convergence theorem, one obtains
\begin{equation} \label{convergence}
\omega_\Delt (A^* \Phi A) \doteq \lim_{n \rightarrow \infty}
\omega_\Delt (A^* \phi(f_n) A) =
\int \! d\rho(\beta) \, \omega_\beta(A^* A) \, \omega_\beta(\phi(0)).
\end{equation}
This relation implies that 
the (according to (\ref{continuity}) continuous) function
\begin{equation} \label{thermfunctions}
\beta \mapsto \Phi (\beta) \doteq \omega_\beta (\phi(0)),
\end{equation} 
called thermal function, determines the central decomposition 
of $\Phi$; hence we may identify $\Phi$ with this function. 
Such central operators are 
called macro--observables in the following. Since 
$\omega_\Delt (\phi(x)) = \omega_\Delt (\phi(0)) = 
\omega_\Delt (\Phi) $, the 
expectation values of the local observables 
$\phi(x)$ in the thermal reference states can be 
interpreted in terms of the macro--observables $\Phi$, as claimed. 

Let us now turn to the characterisation of the local equilibrium states.
Fixing some suitable subspace ${\cal S}_x \subset {\cal Q}_x$,
we say a state $\omega$ is ${\cal S}_x$-compatible with 
a thermal interpretation at $x$ (${\cal S}_x$-thermal, for short)
if there exists some $\omega_{\Delt} \in \CU$  such that  
\begin{equation} \label{compare}
\omega (\phi (x)) = \omega_{\Delt} (\phi (x)), \quad \phi (x) 
\in {\cal S}_x.
\end{equation}
Thus $\omega$ cannot be distinguished from the thermal 
reference state $\omega_\Delt$
by the local observables in ${\cal S}_x$. One can therefore 
consistently assign to it the thermal properties of $\omega_{\Delt}$
described by the macro--observables $\Phi$
corresponding to $\phi (x) \in {\cal S}_x$. More
explicitly, taking into account that 
$\omega (\phi (x)) = \omega_{\Delt} (\phi (x)) = \omega_{\Delt} (\Phi)$, 
one can lift  
$\omega$ at $x$ to the space of macro--observables, setting 
\begin{equation} \label{extension}
\omega(\Phi)(x) \doteq  \omega (\phi (x)), \quad \phi (x) 
\in {\cal S}_x.
\end{equation}
In this way the state $\omega$ acquires at $x$ a physically
meaningful (partial) interpretation in macroscopic terms.

This definition extends in a straightforward manner 
to states which are locally in equilibrium in  
(open) regions ${\cal O} \subset \RR^4$. One first identifies the 
spaces of observables ${\cal S}_x$, $x \in \RR^4$, with the help 
of the automorphic action of the translations, setting 
\begin{equation}
{\cal S}_x \doteq \alpha_x ( {\cal S}_0).
\end{equation}
This being understood, a state $\omega$ is said to be 
${\cal S}_{\cal O}$--thermal if for each $x \in {\cal O}$ there is 
some thermal reference state $\omega_{\Delt_x} \in \CU$ 
such that relation (\ref{compare}) holds. The functions 
\begin{equation}
x \mapsto \omega (\Phi) (x) = \omega (\phi(x)), \quad \phi(0) \in {\cal S}_0,
\end{equation}
then describe the space-time behaviour of the mean values of 
the macro--observables $\Phi$ in the state $\omega$. Hence they provide 
a link between the microscopic dynamics and the evolution of 
macroscopic thermal properties, \ie the thermodynamics of the state. 
It is this aspect which will be studied in the framework of our 
model.

In order to exclude certain pathologies, we restrict
attention here to ${\cal S}_{\cal O}$--thermal 
states $\omega$ admitting reference 
states $\omega_{\Delt_x} \in \CU$, $x \in {\cal O}$, which are weakly 
integrable in $x$ and have temperature supports $\Delt_x$ contained   
in some compact set $\Delt \subset V_+$ for $x$ varying 
in any given compact subset of ${\cal O}$. So the functions
$x \mapsto \omega (\Phi) (x)$ are differentiable in 
the sense of distributions. As a matter of fact, for any 
test function $f$ having compact support in ${\cal O}$
\begin{equation} \label{lift}
\Big|\int \! dx \, f(x) \, \omega (\Phi) (x) \Big|
\leq {\| f \|}_1 \, {\| \Phi \|}_\Delt,
\end{equation}
where ${\| f \|}_1$ denotes the 
$L^1$--norm of $f$, and we have introduced on the thermal functions 
(\ref{thermfunctions}) the seminorms 
\begin{equation} \label{seminorm}
{\|\Phi\|}_\Delt \doteq \sup_{\beta \in \Delt} |\Phi(\beta)|.
\end{equation}
This statement follows from relations (\ref{mixture}),
(\ref{thermfunctions}) and (\ref{compare}).
It will allow us to extend the lifts of local equilibrium 
states to larger spaces of macro--observables.
 
\section{The model}
\setcounter{equation}{0}

In the remainder of this article we apply the preceding
general scheme to the theory of a free massless scalar field. 
After a brief synopsis of the model, we proceed according 
to the steps outlined in the introduction and determine 
the thermal reference states as well as a distinguished space  
of local observables needed for the identification and 
macroscopic interpretation of the local equilibrium states. An
example of such a state is given at the end of this
section.

The free massless scalar field $\phi_0 (x)$ on $\RR^4$
is characterized by its field equation and commutation 
relation  
\begin{equation} \label{freefield}
\square_x \phi_0 (x) = 0, \quad \
[\phi_0 (x_1),\phi_0 (x_2)] = (2 \pi)^{-3} \int \! dp \, e^{-i(x_1-x_2)p}
\, \varepsilon (p_{0}) \delta (p^{2}) \cdot 1,
\end{equation}
which are to be understood in the sense of distributions.
It generates, together with its normal products with reference to the 
vacuum state (\cf the examples below), a polynomial *-algebra ${\cal A}$.
This algebra is stable under the actions of the Poincar\'e group 
${\cal P}_+^\uparrow$,
given by $\alpha_{\Lambda,a} (\phi_0 (x)) = \phi_0 (\Lambda x + a)$
and the gauge group $\ZZ^2$, given by
$\gamma (\phi_0 (x)) = - \phi_0 (x)$.

We restrict attention here to states $\omega$ on ${\cal A}$ which
are gauge invariant, \ie $\omega \circ \gamma = \omega$, so the 
respective $n$-point functions of $\phi_0$ vanish if 
$n$ is odd. The simplest examples of this kind are quasifree states.
They are completely determined by their two-point functions 
through the formula
\begin{equation} \label{wick}
\begin{split}
& {\omega}({\phi_0} (x_{1}){\phi_0}(x_{2}) \cdots {\phi_0}(x_{n}))
\\ 
& \doteq
\begin{cases}
\sum_{\mbox{\tiny pairings}} 
{\omega}({\phi_0}(x_{i_1}) \phi_0(x_{i_2})) \cdots 
{\omega} ({\phi_0}(x_{i_{n-1}}){\phi_0}(x_{i_n})) & 
\text{$n$ even,} \\ 
0 & \text{$n$ odd.}
\end{cases}
\end{split}
\end{equation}
It is a well known fact that the algebra ${\cal A}$ has a unique 
gauge  invariant KMS state $\omega_\beta$ for                   
each temperature vector $\beta \in V_+$. This state is 
quasifree, so it is determined by its two-point function
given by 
\begin{equation} \label{freekms}
\omega_\beta (\phi_0 (x_1) \phi_0 (x_2)) =
(2 \pi)^{-3} \int \! dp \,
e^{-i(x_1-x_2)p} \varepsilon (p_{0})\delta (p^{2})
\frac{1}{1-e^{-\beta p}}.
\end{equation}
The KMS states $\omega_\beta$ comply with our continuity
assumption (\ref{continuity}) and 
fix the convex set $\CU$ of thermal reference states 
which enters into our definition of local equilibrium states.

Next, we describe the spaces ${\cal S}_x$
of local observables which will be used to analyze 
the local equilibrium states. They are generated by 
the unit operator $1$, the field $\phi_0 (x)$ and
balanced derivatives of its normal ordered
square. Introducing the multi-index notation 
$\bmu = (\mu_1, \mu_2, \dots \mu_m)$
and setting 
$\partial_\zeta^\sbmu =
\partial_{\zeta_{\mu_1}}^{ } \cdots \partial_{\zeta_{\mu_m}}^{ }$, 
the latter observables are given by 
\begin{equation} \label{balanced}
\theta^\sbmu (x)  \doteq 
\lim_{\zeta \rightarrow \,0} \,
\partial_\zeta^\sbmu \big( \phi_0 (x + \zeta)\phi_0 (x - \zeta) 
- \omega_\infty \big(\phi_0 (x + \zeta)\phi_0 (x - \zeta)\big) \, 1 \big),
\end{equation}
where the limit is approached from spacelike directions $\zeta$
and $\omega_\infty$ denotes the vacuum 
(\ie $\beta \, {\mbox{\footnotesize $\nearrow$}} \, \infty$) state. 
We assume that all states of interest here are in the domains
of these forms and note that $\theta^\sbmu (x) = 0$
if $m$ is odd since $\phi_0 (x + \zeta)\phi_0 (x - \zeta)$ is even 
in $\zeta$ as a consequence of locality.

As has been explained in Sec.\ 2, each $\theta^\sbmu (x)$
determines some macro--observable $\Theta^\sbmu$.
The corresponding thermal functions have been computed
in \cite{BuOjRo} and are given by 
\begin{equation} \label{parameters}
\beta \mapsto \Theta^\sbmu (\beta) = 
\omega_\beta (\theta^\sbmu (0) )
= c_m \,\partial_\beta^\sbmu \, (\beta^{2})^{-1},
\end{equation}
where $c_m = 0$ if $m$ is odd and 
$c_m = (-1)^{m/2} (4\pi)^m (m +2)!^{-1} B_{m + 2} $ if 
$m$ is even, $B_n$ being the (modulus of the) Bernoulli numbers. 
As $\, \square_\beta \, (\beta^2)^{-1} = 0$ on $V_+$,
the observables $\Theta^\sbmu$ generate only a subspace 
of the space of all macro--observables.
Yet, as we will see, the linear combinations of the 
corresponding thermal functions
are dense in the space of smooth solutions of the wave equation 
on $V_+$ with respect to the topology induced by the seminorms
(\ref{seminorm}). This fact is of relevance since
it implies that many macro--observables of interest, 
such as the entropy density \cite{BuOjRo} or the particle 
density, can be approximated in the states considered here
with arbitrary precision by linear combinations of the $\Theta^\sbmu$. 

\begin{lemma} \label{lemma3.1}
Let $\Xi$ be any smooth solution of the wave
equation on $V_+$, let 
$B \subset V_+$ be compact and let $\varepsilon > 0$.
There are finitely many constants $c_\sbmu$, $c^\prime_\sbmu$ such that 
$$ {\| \, \Xi - {\sum}  c_\sbmu \, \Theta^{\sbmu} \, \|}_\Delt 
< \varepsilon \quad \mbox{and} \quad
{\| \, \partial^\snu \, \Xi - {\sum} c^\prime_\sbmu 
\, \Theta^{\snu \sbmu} \, \|}_\Delt < \varepsilon,$$
\end{lemma}
where $\nu \bmu$ denotes the multi index $(\nu, \mu_1, \mu_2, \dots
\mu_m)$, $\nu = 0,1,2,3$. 

\begin{proof} It is sufficient to establish the statement 
for closed double cones $\Delt$ of the form
$\Delt = \{(\kappa,{\bf 0}) + \overline{V}_{\! +}\}
\cap\{(\kappa^{-1},{\bf 0}) -  \overline{V}_{\! +}\}$
for $0 < \kappa <1$. These regions are stable under the involution
$\iota(\beta) \doteq \beta / \beta^2$, $\beta \in V_+$. As is
well known, this transformation induces an involution on the 
smooth solutions of the wave equation on $V_+$, given by 
$I(\Xi)(\beta) \doteq \iota(\beta)^2 \, \Xi (\iota(\beta))$.
It is continuous relative to the seminorms  (\ref{seminorm}),
$$
\| I(\Xi) \|_\Delt \leq \kappa^{-2} \, \| \Xi \|_\Delt,
\quad \
\| \partial^\nu I(\Xi) \|_\Delt \leq
2 \kappa^{-3} \, \| \Xi \|_\Delt + 3  \kappa^{-4}
\sum_{\nu^\prime} \, \| \partial^{\nu^\prime} \Xi \|_\Delt.
$$
(Similar bounds can be established also for higher
derivatives by elementary computations.) Since
$I$ is an involution, $\Xi$ and $I(\Xi)$ may be interchanged
in these inequalities. 

For the proof of the first half of the statement 
it is convenient to proceed, in a first step, to the functions  
$l_\sbmu \Theta^{\sbmu}$, 
where $l_\sbmu = l_{\mu_1} \dots l_{\mu_m}$, $l$  
being any positive lightlike vector, and, in a second step, 
to their images $I(l_\sbmu  \Theta^{\sbmu})$
under the involution $I$. 
According to the preceding 
remarks it suffices to show that the linear span of 
the latter functions is dense in the space of all 
smooth solutions of the wave equation with regard to 
the seminorms $\|  \cdot  \|_\Delt$.  
The functions $I(l_\sbmu \, \Theta^{\sbmu})$,
resulting from (\ref{parameters}), have the simple form
$$\beta \mapsto I(l_\sbmu  \Theta^{\sbmu}) (\beta) = 
c^\prime_m \, (l\beta)^m,$$
where $c^\prime_m \neq 0$ if $m$ is 
even and $c^\prime_m = 0$ if  $m$ is odd. So we have
to show that suitable linear combinations of the functions 
$\beta \mapsto  (l\beta)^m$ for even $m$ and
lightlike $l$ approximate any given smooth solution of 
the wave equation, uniformly on compact sets $\Delt \subset V_+$.

We make use of the fact  
that the functions $z \mapsto e^{\, \pm \, i \sqrt{z}}$ are analytic in
the complex half--plane 
$\CC_+ \doteq \{z \in \CC : \mbox{Re} \, z > 0 \}$. Thus they can 
be represented on any compact subset $C \subset \CC_+$
by a uniformly convergent power series in $z$, whence 
$z \mapsto e^{\, \pm \, i z} =  e^{\, \pm \, i \sqrt{z^2}}$
can be represented by a uniformly convergent power series
in $z^2 \in C$.
As $\kappa |\bl| \leq l\beta \leq 2 \kappa^{-1} |\bl|$
for $\beta \in \Delt$ and $l = (|\bl|, \bl)$, 
we see that the functions
$\beta \mapsto  e^{\, \pm \, i \,  l\beta} $
can be represented by uniformly convergent power series 
involving only even powers $ (l\beta)^m$, $\beta \in \Delt$. 
Moreover, the functions
$$ \beta \mapsto \sum_{\pm} \int \! d^{\,3}{\bl} \, f_\pm (\bl) \,  
e^{\, \pm \, i \,  l\beta}, $$
where $f_\pm$ are absolutely 
integrable, can be approximated by linear combinations 
of $\beta \mapsto  e^{\, \pm \, i \,  l\beta} $,
uniformly on compact sets $\Delt$. So we conclude that linear 
combinations of $\beta \mapsto  (l\beta)^m$ for even $m$
and suitable $l$ approximate these functions as well.

It remains to show that the restriction of any smooth 
solution $\Xi$ of the wave equation to the double cone $\Delt$
admits an integral representation as given above.
This can be seen if one extends the smooth Cauchy data
of $\Xi$ on the base of $\Delt$
to test functions with compact support in a slightly larger
region. Using
these data as initial values, one obtains a smooth
solution $\Xi_\Delt$ of the wave equation which coincides
with $\Xi$ on $\Delt$ according to standard uniqueness results.  
In view of the regularity properties of its Cauchy
data, $\Xi_\Delt$ admits an integral representation
as given above. In fact, the corresponding
functions $f_\pm$ are, apart from an integrable 
singularity at the origin, smooth and rapidly decreasing.
This completes the proof of the first part of the 
statement.

The proof of the second part proceeds along similar lines
and we therefore indicate only the essential steps. We 
begin by noting that 
$$ \beta \mapsto l_\sbmu  \Theta^{\snu \sbmu} (\beta) = 
 \partial^\snu \, c^\prime_{m+1} \, (l\beta)^m (\beta^2){}^{-m-1},
$$
where $c^\prime_{m+1} \neq 0$ if $m$ is odd. Applying the 
involution $I$ to 
$\beta \mapsto  (l\beta)^m (\beta^2){}^{-m-1}$, 
we see that it suffices to show that suitable linear combinations of 
the functions $\beta \mapsto  (l\beta)^m$ for odd $m$ and
lightlike $l$ approximate any given smooth solution of 
the wave equation, uniformly on compact sets $\Delt \subset V_+$,
and that the same holds true for their respective partial derivatives.

We now use the fact that the functions
$z \mapsto  ({\scriptstyle 1/\sqrt{z}}) \, e^{\, \pm \, i \sqrt{z}}$
are analytic on $\CC_+$. Hence 
$z \mapsto e^{\, \pm \, i z} = ({\scriptstyle z / \sqrt{z^2}}) 
\, e^{\, \pm \, i \sqrt{z^2}}$ 
can be represented, on the domains specified above, by uniformly convergent 
power series involving only odd powers of $z$. So the functions
$\beta \mapsto  e^{\, \pm \, i \,  l\beta} $
can likewise be represented by uniformly convergent power series 
involving only odd powers $ (l\beta)^m$, $\beta \in \Delt$. 
In view of the analyticity of these functions, this uniformity of
convergence holds also for their respective partial derivatives.

Now if the functions $f_\pm$ in the above integral representation
of solutions of the wave equation 
are sufficiently rapidly decreasing, these integrals can not
only be approximated by suitable linear combinations of the 
functions $\beta \mapsto  e^{\, \pm \, i \,  l\beta} $, uniformly
on $\Delt$, but the same holds true also for their respective 
partial derivatives. Yet, as has been explained, the
restrictions of smooth solutions of the wave equation to 
$B$ can be represented in this desired way, so the proof of the 
statement is complete. \end{proof}

In order to illustrate the significance of this result, 
let us consider the mean density of particles of momentum $\bp$.
From a macroscopic point of view, the corresponding observable
consists, for a system in a finite box, of the product of the
annihilation and creation operators of a particle of momentum
$\bp$, divided by the volume of the box. Being in a 
setting describing the thermodynamic limit, this observable is obtained
as limit of sequences of elements of ${\cal A}$
which are bilinear in the basic field $\phi_0$, 
\begin{equation} \label{density}
N_p \doteq \lim_n \, 2 |\bp| \, \phi_0 (f_{p,n})^* \phi_0 (f_{p,n}).
\end{equation} 
Here $p = (|\bp|, \bp)$ and 
$f_{p,n}(x) \doteq n^{-5/2} \, g(n^{-1}x_0) \, h(n^{-1} \bx)
\, e^{\, ipx}$, where $g$, $h$ are test functions satisfying 
$\int \! dx_0 \, g(x_0) = 1$ and
$\int \! d^{\, 3}\bx \, |h(\bx)|^2 = 1$.
The limit exists in all thermal reference 
states and its expectation values can be centrally decomposed 
similarly to relation (\ref{convergence}). Thus each $N_p$
is a macro--observable commuting with all elements of
${\cal A}$. The corresponding thermal function is, as
expected, given by Planck's famous formula for 
the density of massless particles of momentum 
$\bp$ in a thermal equilibrium state,
\begin{equation}
\beta \mapsto N_p(\beta) = (2\pi)^{-3} \, 
\frac{1}{e^{\, \beta p} - 1}.
\end{equation}
As these functions are smooth solutions of the wave equation
on $V_+$, we infer from the preceding lemma that, instead of
relying on $(\ref{density})$, one can determine the particle 
density $N_p$ also locally. More precisely, for given 
temperature support $\Delt$ there 
are local observables in ${\cal S}_x$
which allow to determine the mean values of $N_p$
in all thermal reference states in ${\cal C}_\Delt$ 
with arbitrary precision. Since
all powers of $N_p$ are represented by 
smooth solutions of the wave equation
as well, the same holds true also for all moments of $N_p$
in these states.

Having thus determined all relevant quantities
for the examination of local equilibrium states,
let us turn now to their actual analysis. We will consider
states $\omega$ which are ${\cal S}_{\cal O}$--thermal
in regions ${\cal O}$ of Minkowski space, where 
${\cal S}_x$, \mbox{$x \in {\cal O}$}, are the spaces of local observables
defined above. Taking into account the 
preceding lemma and relation (\ref{lift}), 
these states can be lifted at each 
$x \in {\cal O}$ to all macro--observables
$\Xi$ whose thermal functions are smooth solutions of 
the wave equation on $V_+$. These observables will
be called admissible in the following. We are interested in the 
space--time behaviour of their respective expectation values,
\begin{equation}
x \mapsto \omega(\Xi)(x), \quad \ x \in {\cal O},
\end{equation}
which describe the evolution of these states from a
macroscopic point of view.

An interesting example of a local equilibrium state 
which is ${\cal S}_{V_+}$--thermal was presented
in \cite{BuOjRo}. This state is quasifree and therefore fixed by its 
two--point function which, for $x_1,x_2 \in V_+$, has the form
\begin{equation} \label{bhb}
\omega_{\mbox{\tiny bhb}} (\phi_0 (x_1) \phi_0 (x_2)) \doteq
(2 \pi)^{-3} \int \! dp \,
e^{-i(x_1-x_2)p}\varepsilon (p_{0})\delta (p^{2})
\frac{1}{1-e^{- \eta (x_1 + x_2)p}},
\end{equation}
where $\eta > 0$ is some parameter. As was shown in 
\cite{BuOjRo}, the restriction of $\omega_{\mbox{\tiny bhb}}$ to  
${\cal S}_x$ coincides with the KMS--state 
$\omega_{\beta(x)}$, where $\beta(x) = 2 \eta x$, 
$x \in V_+$. So one obtains for the expectation values of 
the admissible macro--observables $\Xi$ 
\begin{equation}
\omega_{\mbox{\tiny bhb}} (\Xi) (x) = \Xi (2 \eta x), \quad \
x \in V_+,
\end{equation}
where $\beta \mapsto \Xi(\beta)$ is the thermal function
corresponding to $\Xi$.
Plugging the observable $T^2$
determining the square of the temperature 
into this equation (which is legitimate since 
$\beta \mapsto T^2(\beta) = (\beta^2)^{-1}$ is a smooth
solution of the wave equation), one finds that the temperature
of the state $\omega_{\mbox{\tiny bhb}}$
tends to infinity at the boundary of $V_+$. 
As a matter of fact, the state $\omega_{\mbox{\tiny bhb}}$
describes the space--time evolution of a hot bang
at the origin of Minkowski space. 
This can be seen most clearly by looking at the
expectation values of the density operators $N_p$, 
\begin{equation}
\omega_{\mbox{\tiny bhb}} (N_p) (x)
= (2\pi)^{-3} \, 
\frac{1}{e^{\, 2 \eta x p} - 1}, \quad \ x \in V_+.
\end{equation}
If $x$ approaches the boundary of the lightcone $V_+$,
this expression stays bounded unless $x$ and the four--momentum $p$
become parallel. Thus the bulk of the particles 
giving rise to the infinite temperature at the boundary
of the lightcone originates from its apex.
Whence the state $\omega_{\mbox{\tiny bhb}}$ 
provides a qualitative picture of the space--time patterns of
the radiation caused by a hot bang. We will see that the 
occurrence of such singularities is a generic feature of 
the present model.

\section{Evolution equations}
\setcounter{equation}{0}

We want to demonstrate now how 
the microscopic dynamics leads to 
evolution equations for the expectation values of 
macro--observables in local equilibrium states
in the present simple model.
We begin by noting that the field equation for $\phi_0$ 
entails the following equations (in the sense of distributions)
\begin{eqnarray}
& \partial_x{}_{\snu} \, \partial_{\zeta}{}^\snu \, \phi_0(x + \zeta) 
\phi_0(x-\zeta) = 0, & \\
& \square_x \, \phi_0(x + \zeta) \phi_0(x-\zeta) = 
- \, \square_\zeta \, \phi_0(x + \zeta) \phi_0(x-\zeta). &
\end{eqnarray}
This implies for the local observables
$\theta^\sbmu$, introduced in relation (\ref{balanced}),
\begin{eqnarray}
& \partial{}_{\snu} \, \theta^{\snu \sbmu}(x) = 0, & \label{4.3} \\
& \square  \, \theta^{\sbmu}(x)  = - \, 
\theta_\snu^{\snu \sbmu}(x). & \label{4.4}
\end{eqnarray}
In particular, each $\theta^{\sbmu}$ is a conserved tensorial
current. The latter relations lead to linear evolution
equations for the expectation values of the corresponding 
macro--observables $\Theta^\sbmu$ in local equilibrium
states. As a matter of fact, 
the following lemma holds.
\begin{lemma} \label{lemma4.1}
Let $\omega$ be an ${\cal S}_{\cal O}$--thermal
state and let $\Xi$ be any admissible macro--observable. Then
$$
\partial{}_{\snu} \, \omega(\partial^{\, \snu} \Xi)(x) = 0, \quad \
\square  \, \omega (\Xi) (x)  = 0 \quad \ \mbox{for} \ x \in {\cal O},
$$
where $\partial^{\, \snu} \Xi$ is the 
macro--observable with thermal function 
$\beta \mapsto \partial^{\, \snu} \,  \Xi(\beta)$.
\end{lemma}

\begin{proof} For the proof of the first equation we proceed from 
relation (\ref{4.3}) which implies for any choice of constants
$c_\sbmu$
$$  \partial{}_{\snu} \, \omega\big({\sum} c_\sbmu 
\, \theta^{\snu \sbmu}(x)\big) = 0.
$$
As $\omega$ is ${\cal S}_{\cal O}$--thermal, we also have 
for $x \in {\cal O}$
$$
\omega\big({\sum} c_\sbmu \, \Theta^{\snu \sbmu}\big)(x) =
\omega\big({\sum} c_\sbmu \, \theta^{\snu \sbmu}(x)\big),
$$
and combining these two equations we arrive at 
$$  \partial{}_{\snu} \, \omega\big({\sum} c_\sbmu 
\, \Theta^{\snu \sbmu}\big)(x) = 0.
$$
The first part of the statement now follows from the continuity
property (\ref{lift}) of local equilibrium states and Lemma \ref{lemma3.1}.

For the second equality we make use 
of relation (\ref{4.4}) which implies
$$
\beta \mapsto 
\Theta_\snu^{\snu \sbmu}(\beta) = 
\omega_\beta(\theta_\snu^{\snu \sbmu}(x))
= - \, \omega_\beta(\square  \, \theta^{\sbmu}(x))
= - \, \square  \, \omega_\beta(\theta^{\sbmu}(x)) = 0,
$$
where the latter equality follows from the invariance of the 
KMS states under space--time translations. Hence 
$\Theta_\snu^{\snu \sbmu} = 0$, so taking into account once more that
$\omega$ is ${\cal S}_{\cal O}$--thermal we obtain
\begin{gather*}
\square  \, \omega\big(\, {\sum} c_\sbmu \Theta^{\sbmu}\big)(x)
= \square  \, \omega\big(\, {\sum} c_\sbmu \theta^{\sbmu}(x)\big)
= \omega(\, {\sum} c_\sbmu \, \square  \, \theta^{\sbmu}(x)) \\
 = - \, \omega\big(\, {\sum} c_\sbmu \theta_\snu^{\snu \sbmu}(x)\big)
= - \, \omega\big(\, {\sum} c_\sbmu \Theta_\snu^{\snu \sbmu}\big)(x) = 0. 
\end{gather*}
As before, the second assertion then follows from the continuity properties
of $\omega$ and Lemma \ref{lemma3.1}.
\end{proof}

A macro--observable of particular physical interest is the particle
density $N_p$. Its evolution in local equilibrium states is described
in the following proposition.

\begin{prop}  Let $\omega$ be any ${\cal S}_{\cal O}$--thermal state
and let $p$ be any positive lightlike vector. Then 
$$
(p \, \partial) \, \omega(N_p) (x)= 0, \quad \
\square \,  \omega(N_p) (x)= 0 \quad
\quad \mbox{for} \ x \in {\cal O}.
$$
\end{prop}

\begin{proof} Consider the function $L_p$ on $V_+$,
$$
\beta \mapsto L_p(\beta) \doteq (2\pi)^{-3} \, \ln \, (1 -e^{\, - \beta p}).
$$
It is a smooth solution of the wave equation and consequently
$L_p$ is an admissible macro--observable. Now
$$
\partial^{\, \snu} L_p(\beta) = p^{\snu} \, 
 (2\pi)^{-3} \, \frac{1}{e^{\, \beta p} - 1}
=  p^{\snu} N_p(\beta).
$$
Applying the preceding lemma, we conclude that
$$
(p \, \partial) \, \omega(N_p) (x) =
\partial_\snu \, \omega(p^\snu N_p) (x)
= \partial_\snu \, \omega(\partial^\snu L_p) (x) = 0,
$$
as claimed. The second part of the statement follows
directly from the lemma.~\end{proof}

The preceding result shows that the local 
equilibrium situations, described 
by ${\cal S}_{\cal O}$--thermal 
states, are compatible with those considered in 
the context of transport equations, such as 
the Boltzmann equation for the particle 
distribution function $x,p \mapsto \overline{N}(x,p)$.
In the latter case one characterizes the 
distribution functions $\overline{N}$ describing 
local equilibrium by the condition that the collision 
term in the Boltzmann equation vanishes. 
In the present approach these distributions are
to be identified with $x,p \mapsto \omega(N_p) (x)$, and
the first part of the proposition shows that 
these functions satisfy the collisionless Boltzmann equation
as well. From the second part
we see that they have to comply with  
an additional constraint of dynamical origin.

We mention as an aside that the functions 
$x,p \mapsto \overline{N}(x,p)$ 
contain all macroscopic information about 
the local equilibrium states. 
As a matter of fact, given any non--negative function 
$\overline{N}$ satisfying in some convex space--time region
${\cal O}$ the two equations in the proposition
and being sufficiently well--behaved in $p$, one can define
a quasifree functional $\varphi$ on the  
algebra ${\cal A}$ of local observables, setting for $x_1,x_2 \in {\cal O}$
\begin{alignat}{1}
& \varphi(\phi_0(x_1) \phi_0(x_2))  \nonumber \\
\doteq \, & \omega_\infty(\phi_0(x_1) \phi_0(x_2))  
+ \int \! d^{\, 3}\bp \, |\bp|^{-1} 
\, \overline{N}\big((x_1 + x_2)/{2},p\big) 
\, \cos\big((x_1 - x_2)p \big).
\end{alignat}
This functional is real and satisfies all linear constraints imposed
by the field $\phi_0$, but it need not be positive. Yet if it complies
with the latter condition, it 
defines a local equilibrium state in the region ${\cal  O}$.
As we shall see in the subsequent section, the condition of  
positivity imposes further stringent constraints.

\section{Regions of local equilibrium}
\setcounter{equation}{0}

In this section we want to determine 
the shape of space--time regions in which local 
equilibrium is possible. We restrict attention here to 
convex regions. Given 
a state $\omega$, let ${{\cal O}_\omega}$ be
any maximal (in-extendible) region of this kind
in which $\omega$ is ${\cal S}_{{\cal O}_\omega}$--thermal.
As was exemplified in Sec.~3, there are states  
where ${{\cal O}_\omega}$ contains a 
lightcone. It will turn out that, conversely,   
any such ${{\cal O}_\omega}$ has to be contained in some
timelike cone unless the lifts of $\omega$ to the 
admissible macro--observables coincide at all
points of ${{\cal O}_\omega}$
with some fixed thermal reference state.

It will be convenient in this analysis to proceed from 
$\omega$ to suitably regularized states.
To this end we pick some non--negative test function $f$ 
which has support in a small neighbourhood of 
$0 \in \RR^4$ and satisfies $\int \! dy \, f(y) = 1$.
The corresponding regularized state is given by 
$\omega_f \doteq \int \! dy \, f(y) \, \omega \circ \alpha_y$,
hence $\omega_f \circ \alpha_x = \omega_{f_{-x}}$.
As the set  ${\cal C}$ 
of thermal reference~states is convex, $\omega_f$
is ${\cal S}_{{\cal O}}$--thermal in some slightly smaller 
region ${\cal O} \subset {\cal O}_\omega$. 
Thus, taking into account 
relation (\ref{lift}) and Lemma \ref{lemma3.1},
we obtain for all admissible macro--observables $\Xi$
and multi indices $\bmu$ the bounds
\begin{equation} \label{5.1}
|\, \partial^{\,\sbmu}  \omega_f (\Xi) (x) \, | \leq c^{\,\sbmu} \, 
\| \Xi \|_\Delt,
\end{equation}
uniformly on compact subsets of ${\cal O}$, where 
$c^{\,\sbmu}$ are certain constants and 
$\Delt \subset V_+$ is compact.

We will analyze the states $\omega_f$ on subspaces 
$\boldsymbol{\Gamma}_p$
of macro--observables, fixed by the positive lightlike vectors 
$p$. These spaces consist of operators whose thermal functions 
are of the form $V_+ \ni \beta \mapsto \Gamma(\beta p)$, 
where $\Gamma$ is smooth on $\RR_+$. 
So they are admissible macro--observables. 
As their sums, products and adjoints 
are also of this form, the 
spaces  $\boldsymbol{\Gamma}_p$ are
in fact abelian $*$--algebras. Of special interest are their 
elements $E_p$ and $M_p^{\, n}$ given by 
\begin{equation} \label{obsdef}
\beta \mapsto E_p(\beta) \doteq e^{\, i \, \beta p}, \quad \quad
\beta \mapsto M_p^{\, n}(\beta) \doteq (\beta p)^n, \ n \in \NN_0.
\end{equation}
Since $\omega_f$ is a state, its lift $\omega_f(\,\cdot\,)(x)$
to the admissible 
macro--observables at any point $x \in {\cal O}$ 
defines a state on each $\boldsymbol{\Gamma}_p$. We collect
some pertinent properties of these lifts 
in the subsequent lemma.
\begin{lemma} Let  $\omega_f$ be any regularized 
${\cal S}_{\cal O}$--thermal state. Then, for $x \in {\cal O}$, 
\vspace*{-2mm}
\begin{itemize}
\item[(a)] $(p \, \partial) \, \omega_f(E_p)(x) = 0, \quad
     \square \, \omega_f(E_p)(x) = 0,$  
\item[(b)] $| \, \omega_f(E_p)(x) \, | \leq 1$,  
\item[(c)] $p \mapsto \omega_f(E_p)(x)$ extends to  
an entire analytic function on $\CC^4$ which is smooth in $x$,  
\item[(d)] 
$0 \leq {\omega_f(M_p^{\, n + 1})(x)} \leq \big(\omega_f(M_p^{\,
  n})(x)\big)^{n/(n+1)} \,  
\big(\omega_f(M_p^{\, 2n + 1})(x)\big)^{1/(n+1)} $, 
\ $n \in \NN_0$.
\end{itemize}
\end{lemma}

\begin{proof} Statement (a) follows
from Lemma  (\ref{lemma4.1}) and the relation
$p^\snu E_p = -i \, \partial^\snu E_p$. For the proof of (b)
we make use of the fact that
$\omega_f(\,\cdot\,)(x)$ is a state on  $\boldsymbol{\Gamma}_p$,
hence $|\, \omega_f(E_p)(x) \,|^2 \leq 
\omega_f(E_p{}^* E_p)(x) = 1$, $x \in {\cal O}$. 
In order to establish the existence of the extension 
in (c), we first extend the operators $E_p$ 
to admissible operators $E_k$, $k \in \CC^4$. Their respective
thermal functions are given in proper coordinates by 
$$
\beta \mapsto 
E_k(\beta) \doteq \big(\cos(\beta_0|\bk|) + ik_0 \, |\bk|^{-1}
\sin(\beta_0|\bk|) \big) \, e^{\, -i \sbk \sbbeta}.
$$
These functions are entire analytic in $k$ and 
locally uniformly bounded if $\beta$ varies in compact sets
$\Delt \subset V_+$. It thus follows from 
the continuity properties subsumed in (\ref{5.1})
that $ k \mapsto \omega_f(E_k) (x)$ provides the desired extension.
The remaining statement (d) is a straightforward consequence of 
H\"older's inequality, taking into account that the 
operators $M_p^{\, n}$, $n \in \NN_0$, are positive elements of
$\boldsymbol{\Gamma}_p$ (being represented by positive 
functions on $V_+$). \end{proof}

After these preparations we can establish now the limitations 
on the possible shape of regions of local equilibrium, indicated
above.
\begin{prop} Let $\omega$ be a state which is 
${\cal S}_{{\cal O}_\omega}$--thermal in some 
convex region ${{\cal O}_\omega}$ containing 
a lightcone. There are the following alternatives
for the lifts $\omega(\, \cdot \,)(x)$
of $\omega$ to the admissible macro--observables:

\vspace*{-2mm}
\begin{itemize}
\item[(a)] The lifts do not depend on $x \in {\cal O}_\omega$, \ie
they all coincide with some fixed thermal reference state.
\item[(b)] The lifts depend non--trivially on $x \in {\cal O}_\omega$, 
and there is some timelike simplicial cone (an intersection of characteristic
half spaces) containing  ${\cal O}_\omega$. 
\end{itemize}
\end{prop}

\begin{proof} 
As explained above, we proceed from $\omega$ to the   
regularized states $\omega_f$.
Any one of the latter states is  ${\cal S}_{\cal O}$--thermal in the 
region ${\cal O} = \bigcap_{\, x \, \in \, \mbox{\footnotesize supp} \, f} 
\{ {\cal O}_\omega + x \}$, which likewise is convex and contains some
lightcone, say $V_+$ for concreteness.
Because of the convexity
of ${\cal O}$, $V_+ + y \subset {\cal O}$ for $y \in {\cal O}$. 
We consider now the functions, $p$ being an arbitrary 
positive lightlike vector, 
$$
x \mapsto \overline{E}_p(x) \doteq \omega_f (E_p)(x), \quad x \in {\cal O}.
$$
Introducing the coordinates $x_\pm = x_0 \, \pm \, \be \bx$ 
and $\bx_\bot = \bx - (\be \bx) \be$, where $\be = \bp/|\bp|$,
it follows from part (a) of the preceding lemma that $\overline{E}_p$
does not depend on $x_+$ and satisfies $\triangle_\bot \overline{E} _p(x) = 0$,
$x \in {\cal O}$, where $\triangle_\bot$ is the Laplacian with
respect to $\bx_\bot$. 

Now, given $x_- \in \{y_-: y \in {\cal O} \}$ and any 
$\bx_\bot \in \RR^2$, there exists an $x \in {\cal O}$ with these 
components. For if $y \in {\cal O}$, the point $x$ with the 
components $x_- = y_-$, $x_+ = y_+ + t$, $\bx_\bot$ is,
for sufficiently large $t$, contained
in the lightcone $z + V_+ \subset {\cal O}$, provided  
$z \in {\cal O}$ and $(y -z)$ is positive timelike.
So for fixed $x_- \in \{y_-: y \in {\cal O} \}$ one has  
$\triangle_\bot \overline{E} _p(x) = 0$, $\bx_\bot \in \RR^2$.
But according to parts (b) and (c) of the preceding lemma,
$x \mapsto  \overline{E}_p(x)$ is smooth and bounded
in modulus by $1$, hence it cannot depend on
$\bx_\bot$ in view of the growth properties of 
non--trivial solutions of 
the Laplace equation (Harnack's inequality).

Because of the analyticity properties of $\overline{E}_p$
established in part (c) of the lemma, we can represent
this function as power series,
$ \overline{E}_p(x) = 
\sum_{m = 0}^{\infty} \, |\bp|^m \, c_\sbmu(x) \, e^{\sbmu},
$
where $e = (1,\be)$ and all coefficients $c_\sbmu(x)$ are smooth in $x$. 
Moreover, in view of the results obtained in the 
preceding step, $x \mapsto c_\sbmu(x) e^\sbmu$  
can depend only on $x_- = ex$ in a 
\mbox{non--trivial} manner. Thus, by 
$k$--fold partial differentiation with respect to $x$ and an
application of the chain rule, we obtain
$$   
(y \, \partial)^{k} \, c_\sbmu(x)  e^\sbmu =
(ye)^{k} \, \partial_0^{\,k} \, c_\sbmu(x) e^\sbmu,
\quad     x \in {\cal O}, \, y \in \RR^4.
$$
Expressing the products of the components of $\be$
in terms of spherical harmonics, it is apparent that, for $k > m$,
this equality can only be satisfied if 
$\partial_0^{\,k} \, c_\sbmu(x)  e^\sbmu = 0$.
So each function 
$x \mapsto c_\sbmu(x) e^\sbmu$
is a polynomial in $ex$ of degree $m$ or less
with coefficients which are polynomials in the components of $\be$.

Making use of the fact that 
$\overline{E}_p(x) = \sum_{m = 0}^{\infty} \, 
\frac{|\sbp|^m \, i^m}{m!} \, \omega_f(M_e^{\,m})(x),
$
where $M_e^{\, m}$ are the operators 
introduced in (\ref{obsdef}), we therefore find that each function
$$
x \mapsto {\omega_f(M_e^{\,m})(x)} = 
(-i)^m m! \ c_\sbmu(x) \, e^\sbmu,
\quad     x \in {\cal O},
$$
is a polynomial in $ex$ of degree 
$m$ or less. According to part (d) of the preceding 
lemma these polynomials are non--negative.
Moreover, there is some $m$ such that the 
corresponding polynomial is of first degree in $ex$, 
unless $x \mapsto {\omega_f(M_e^{\,m})(x)}$
is constant for all $m \in \NN$. To verify this, 
let $x \mapsto {\omega_f(M_e^{\,k})(x)}$
be constant for $k = 0, \dots m-1$. It then follows from
(d) and the fact that the polynomial degree of 
$x \mapsto {\omega_f(M_e^{\,2m - 1})(x)}$ is 
at most $2m -1$ that 
$x \mapsto {\omega_f(M_e^{\,m})(x)}$ increases
no faster than $|x|^{(2m-1)/m}$ for large 
$x \in V_+  \subset {\cal O}$. Hence 
$x \mapsto {\omega_f(M_e^{\,m})(x)}$, being a  polynomial, 
has to be of first degree in $ex$ if it is not constant.

If the second alternative in the statement holds, \ie if
$x \mapsto \omega_f(\Xi)(x)$ depends non--trivially on $x$ for 
some $\Xi$, there is also some $M_e^{\,m}$ for which this
is true since the linear span of the latter operators is dense in the
space of admissible macro--observables, \cf the proof of 
Lemma \ref{lemma3.1}. Thus, because of the preceding 
results, we may assume that 
$$
\omega_f(M_e^{\,m})(x) = P_f(\be) \, (ex) + Q_f(\be), \quad x \in {\cal O},
$$
where $P_f$, $Q_f$ are polynomials and $P_f$ is not identically zero.
But $\omega_f(M_e^{\,m})(x) \geq 0$, $\quad x \in {\cal O}$, so 
$$
P_f(\be) \, (ex) + Q_f(\be) \geq 0, \quad x \in {\cal O}.
$$
Since $V_+ \subset {\cal O}$ and $P_f \neq 0$
it follows that
$P_f(\be) > 0$ for almost all $\be $ and $Q_f(\be) \geq 0$. Hence   
${\cal O} \subset \bigcap_{\, \sbe} \, 
\{ x : ex \geq - Q_f(\be)/P_f(\be) \}$.
In particular, ${\cal O}$ is contained in some positive timelike simplicial 
cone consisting of the intersection of four characteristic half--spaces.
The same is therefore true for ${\cal O}_\omega$ since the support 
of $f$ is compact.

We mention as an aside that, due to the linearity of $\omega_f$ in $f$ and 
$\omega_f \circ \alpha_y = \omega_{f_{-y}}$, one has 
$ P_f(\be) = \int \! dy \, f(y) \, P(\be)$ 
and 
$ Q_f(\be) = \int \! dy \, f(y) \, \big((ey) P(\be) + Q(\be)\big)$
for certain polynomials $P$, $Q$ which do not depend on $f$.

The statement has thus been established for regions ${\cal O}_\omega$
containing $V_+$. But the arguments can be carried over to arbitrary
regions containing some shifted forward or backward lightcone, so the proof
of the proposition is complete. \end{proof}

The preceding result shows that non--trivial local equilibrium 
states $\omega$ fix a time direction in the sense that 
they determine 
some maximal timelike cone in which local equilibrium can exist.
In spite of the fact that the underlying theory is time reflection invariant, 
an observer who has such a state as a background may consistently 
take the corresponding cone as an indicator of the future direction.
If he perturbs $\omega$ by some local 
unitary operation $U$, 
$\omega_U(\, \cdot \,) \doteq \omega(U^* \cdot U)$,
he finds that the perturbed state $\omega_U$ coincides with $\omega$
with regard to all measurements performed after the perturbation,
\ie in the future of the space--time support of $U$. This is a 
consequence of the timelike commutativity of 
observables (Huyghens' principle) in the present model. So, in a sense, 
the model also describes the return to (local) equilibrium. These 
results show that the arrow of time 
manifests itself already in the macroscopic space--time patterns of local
equilibrium states.

Of particular interest is the situation where  
$\omega$ describes a non--trivial macroscopic hydrodynamic flow
in some timelike cone, \ie the local rest systems of $\omega$, which can be 
determined by the admissible macro--observables 
$\beta \mapsto M_e(\beta) \doteq e \beta$, $e$ being positive lightlike, 
vary from point to point. In this case the arguments given in the preceding
proposition imply that the function 
$x \mapsto \omega(M_e)(x)$ has to be of the form 
\begin{equation}
\omega(M_e)(x) = c_\omega \, e(x - x_\omega), 
\end{equation}
where $c_\omega$ is some non--zero (say, positive) constant
and $x_\omega$ some constant vector. 
If an observer finds such a behaviour 
in some bounded space--time region,  
he would infer that the maximal timelike cone 
where $\omega$ can be locally in equilibrium is 
the lightcone $V_+ + x_\omega$ since 
$\omega(M_e)(x)$ has to be non--negative. Anticipating that
$\omega$ has this property, he would also conclude that 
the functions $x \mapsto \omega(M_e)(x)$ vanish at the   
apex $x_\omega$ of this cone, hence the temperature 
of the state tends to infinity there. Moreover,
if $x$ approaches any point on the ray $x_\omega + \RR_+ e$ 
at the boundary of the cone, the time axes of the 
respective local rest systems of the state bend towards the
direction of $e$, indicating that there is a 
dominant flow of particles originating from $x_\omega$.
This can be made more explicit by looking at the 
particle density $x \mapsto \omega(N_p)(x)$ which
can be shown to diverge if $x-x_\omega$ and $p$ become parallel.
So the observer would be led to interpret the underlying 
state as the result of a hot bang which has occurred 
at $x_\omega$ and has 
quickly approached local equilibrium thereafter.
Of course, all his conclusions depend on the hypothesis
that the partial state which he observes is locally in 
equilibrium in all of $V_+ + x_\omega$. Whether this is 
really the case can evidently not 
be decided by observations made in any finite space--time 
region.

\section{Summary and outlook}
\setcounter{equation}{0}

In the present investigation we have applied the general 
formalism for the characterization and analysis of local 
equilibrium states, established in \cite{BuOjRo}, 
to a simple model. There are two elemental ideas underlying 
this approach: (a) local equilibrium states cannot be 
distinguished from global ones by observations made 
in small space--time regions and (b) local measurements 
admit a macroscopic interpretation in all global equilibrium 
states. The combination of these two basic facts made it possible to 
interpret within the microscopic setting 
the space--time patterns of local equilibrium 
states in macroscopic terms.

The results of our investigation show that this approach not 
only is physically meaningful, but it also provides a convenient 
framework for concrete computations. In the case at hand, we 
have been able to establish equations for the space--time 
evolution of macroscopic properties of local equilibrium 
states, such as the particle density, without having to enter 
into the difficult dynamical question of how a system is
driven towards local equilibrium. No {\it a priori} hypotheses,
such as Boltzmann's assumption of molecular chaos, were necessary to
establish these results. Instead, the intrinsic 
characterization (a) 
of local equilibrium was sufficient for their derivation.

The present approach revealed also   
interesting constraints on the possible macroscopic 
space--time structure of local equilibrium states. It
turned out that local equilibrium can exist only in
certain specific space--time regions which generically have
the form of lightcones. Such states can 
therefore be used to distinguish between future and past. 
Moreover, typically there appear singularities at the apices of 
these cones which have the interpretation of a hot bang.
To the best of our knowledge, these results 
provide the first example of
a ``singularity theorem'' in a microscopic setting.

The present model of a massless free field 
exhibits a greatly simplifying feature, however:
it is scale invariant, \ie physics looks the same at 
microscopic and macroscopic scales. For a study of 
the effects of an inherent scale, one may  
consider the example of massive free field 
theory. Indeed, there appear some differences.
It is not difficult to show that, in the massive case,     
non--trivial local equilibrium states having 
locally a sharp temperature do not exist,
their temperature support is typically all of $V_+$. This
suggests to enlarge the space ${\cal C}$ 
of thermal reference states accordingly. In this 
extended framework it is still possible to establish
evolution equations for the thermal equilibrium 
states, in complete analogy to the present results, 
and to analyze their space--time behaviour. These
results will be put on record elsewhere.

It seems attractive to apply, in a further step, the 
general formalism in \cite{BuOjRo} to free fields on arbitrary
globally hyperbolic space--times. In particular,  
the shape of regions admitting local equilibrium
is of interest there as well. For maximally symmetric spaces 
there should appear no new 
conceptual difficulties in such an analysis since the 
basic ingredients, local observables 
and ``global'' equilibrium states for geodesic 
\mbox{observers} (\ie states which are
in equilibrium in the causal closure of the respective 
geodesics), do exist in these cases. But in generic spacetimes 
the situation is less clear in view of the fact that  
such global equilibrium states need not exist.
Thus what is required there is a    
substitute for these states which allows one to identify
the local equilibrium situations.
For the solution of this problem the 
generally covariant framework of quantum field theories,
proposed in \cite{BrFrVe}, 
seems to be of relevance since it allows the unified treatment 
of a given theory on arbitrary spacetimes. So one may
be able to use the existence of global equilibrium states 
on one spacetime to interpret the local equilibrium properties of 
states on the others. 

We mention as an aside that the present
approach to the characterization of equilibrium states 
may be regarded as a refinement of the ideas underlying
the Hadamard condition \cite{Wa} which is of fundamental importance  
in this setting. Yet whereas the Hadamard condition
has been invented to characterize locally the folia of all 
physical states, the 
condition of local equilibrium is designed to identify within
these folia states having a specific physical interpretation. 
Thus the present ideas seem to be of relevance also for the 
interpretation of states on arbitrary space--time manifolds.

Returning to the case of Minkowski space, a central  
issue is the treatment of interacting
models where one may study the intriguing question
of the approach to equilibrium.  
In order to indicate the problems appearing in this context, 
let us consider first a state $\omega$ which is 
locally thermal on a space of observables generated by 
balanced derivatives $\theta^\sbmu(x)$ of 
the normal ordered squares of the underlying fields~\cite{BuOjRo,Zi},
similarly to the observables (\ref{balanced})
used in the present investigation.
In the interacting case, these observables no longer 
satisfy the equations (\ref{4.3}) and (\ref{4.4}),   
which were the basis for the present analysis, 
there appear additional ``source terms'' $\sigma^{\sbmu}(x)$,
$\tau^{\sbmu}(x)$,
\begin{equation} \label{6.1}
\partial{}_{\snu} \, \theta^{\sbnu \sbmu}(x) = \sigma^{\sbmu}(x), \quad
\square  \, \theta^{\sbmu}(x)  =  \tau^{\sbmu}(x). 
\end{equation}
Denoting by $\Theta^{\sbmu}$ the macro--observables
corresponding to $\theta^\sbmu(x)$, one thus obtains 
equations for their evolution in state $\omega$ of the form
\begin{equation} \label{6.2}
\partial{}_{\snu} \, \omega(\Theta^{\sbnu \sbmu})(x) 
= \omega(\sigma^{\sbmu}(x)), \quad
\square  \, \omega(\Theta^{\sbmu})(x)  = \omega(\tau^{\sbmu}(x)). 
\end{equation}
The terms appearing on the right hand side of these equations 
resemble the collision terms in transport equations
\cite{ZuMoRo}. In particular, they vanish if $\omega$
is sufficiently close to equilibrium at $x$ since
$ \sigma^{\sbmu}(x)$, $ \tau^{\sbmu}(x)$ have zero 
expectation values in all thermal reference states due
to their space--time invariance. 

In order to establish the approach to (local) equilibrium one 
thus has to show that these collision terms vanish asymptotically. 
The present results suggest that,
for a given state $\omega$, one may expect such a  
behaviour at best in some lightcone and, 
because of the PCT theorem, this cone may be either future
or past directed. So one has to identify the convex
subsets of states exhibiting the desired behaviour
for given time direction. Mixtures of the two types of 
states may in general not exhibit an ergodic 
behaviour. 

Next, there is the problem of local thermalization of 
a given state $\omega$ on the space of  
observables $\theta^\sbmu(x)$, which was taken for granted
in the preceding discussion. Its solution requires a proof 
that there are asymptotic bounds of the form~\cite{BuOjRo}
\begin{equation} \label{6.3}
\big| \, \omega\big({\sum}  c_\sbmu \, \theta^{\sbmu}(x)\big) \, \big|
\leq {\big\| \, {\sum}  c_\sbmu \, \Theta^{\sbmu} \, \big\|}_\Delt
\, ,
\end{equation}
where $\| \, \cdot \, \|_\Delt$ denotes any one of the seminorms
on the macro--observables, introduced in Sec.\ 2. 
Choosing $\Delt$ sufficiently large, it 
seems feasible to establish such a result without major difficulties 
for finite dimensional spaces of observables $\theta^\sbmu(x)$ whose 
corresponding thermal functions $\Theta^{\sbmu}$ are linearly 
independent. Yet if there exist relations between these
functions of the form ${\sum}  c_\sbmu \, \Theta^{\sbmu} = 0$
which do not have a counterpart at the microscopic
level, one has to understand the mechanism by which 
the expectation values of the corresponding combinations
${\sum}  c_\sbmu \, \theta^{\sbmu}(x)$ of local
observables vanish asymptotically in the state  $\omega$. It is of 
interest in this context that there exists a large family of 
states which may be regarded as perturbations of global equilibrium 
states and which satisfy the above inequality \cite{BuOjRo}.

Once local equilibrium has been reached in a state ${\omega}$, 
either exactly or in the sense that the collision terms in (\ref{6.2}) 
may effectively be neglected, one arrives at the same 
conservation laws and evolution equations  
for the macro--observables 
$\Theta^{\sbmu}$ in the state ${\omega}$ as in free field theory,
\begin{equation} \label{6.4}
\partial{}_{\snu} \, \omega(\Theta^{\sbnu \sbmu})(x) 
= 0, \quad
\square  \, \omega(\Theta^{\sbmu})(x)  = 0. 
\end{equation}
These equations thus have a universal character.
Yet in order to be able to interpret them in physical terms 
it is necessary to determine the underlying thermal functions 
$\beta \mapsto \Theta^{\sbmu} (\beta)$, which are
model dependent.
This requires a computation of the thermal two--point 
functions of the underlying fields. The general form of 
these functions is known \cite{BrBu} and there exist
also various methods for their perturbative computation
in models \cite{LaWe,Ka}, \cf also the novel approach in 
\cite{BrBu2}. It seems therefore possible
to determine the macroscopic space--time patterns 
of local equilibrium states in the presence of interaction
by similar methods as in the present investigation. 
We hope to return to this interesting issue elsewhere.

Let us finally mention that the basic ideas underlying the
approach in \cite{BuOjRo} can also be applied to non--relativistic
theories, such as spin systems. Of course, the concept of local
observable, as used in the present analysis,  
is no longer meaningful there, one has to replace it 
by suitable subspaces of the algebra of all observables.
A general characterization of these subspaces is still an open problem.
But a preliminary study \cite{ArBu} of the $XX$--model indicates that 
certain macroscopic aspects of local equilibrium states, such as 
the existence of 
evolution equations, can still be established in such a setting,
at least in simple cases. So there is evidence 
that the general approach to the analysis of 
local equilibrium states, proposed in \cite{BuOjRo}, provides
an efficient framework for the analysis of this important issue.

\vspace*{3mm}
\noindent {\Large \bf Acknowledgements}\\[2mm]
I would like to thank H.\ Araki, I.\ Ojima and H.\ Roos for
discussions on this topic and I am grateful for hospitality 
and financial support by RIMS, Kyoto University.

\end{document}